\documentstyle[12pt]{article}

\newcommand{\beq}[1]{\begin{equation}\label{#1}}
\newcommand{\eeq}{\end{equation}}

\newcommand{\rref}[1]{(\ref{#1})}
\newcommand{\bqn}{\begin{eqnarray}}
\newcommand{\eqn}{\end{eqnarray}}

\begin{document}

{\flushright{\small Imperial/TP/97-98/56\\DFTUZ 98/\\
    hep-th/9807216\\}}

\begin{center}
  {\Large {\bf Quantum three-dimensional de Sitter space}}
  
  \vspace{0.2in}
  
  Maximo Ba\~nados\footnote{On leave from: Centro de Estudios
    Cient\'{\i}ficos de Santiago, Casilla 16443, Santiago, Chile and,
    Departamento de F\'{\i}sica, Universidad de Santiago, Casilla 307,
    Santiago, Chile.}  \vskip 0.2cm {{\it Departamento de F\'{\i}sica
      Te\'orica, Facultad de Ciencias,\\ Universidad de Zaragoza,
      Zaragoza 50009, Spain}
    \\
    \small E-mail: {\tt max@posta.unizar.es}} \vspace{0.2in}
  
  Thorsten Brotz \vskip 0.2cm {{\it Blackett Laboratory, Imperial
      College of Science, Technology
      and Medicine,\\
      Prince Consort Road, London SW7 2BZ, UK}
    \\
    \small E-mail: {\tt t.brotz@ic.ac.uk}} 
    \\
      and
    \\
    {\it Fakult\"at f\"ur Physik, Universit\"at Freiburg, 
      Hermann-Herder-Strasse 3,\\
      D-79104 Freiburg, Germany}\vspace{0.2in}
  
  Miguel E. Ortiz \vskip 0.2cm {{\it Blackett Laboratory, Imperial
      College of Science, Technology
      and Medicine,\\
      Prince Consort Road, London SW7 2BZ, UK}
    \\
    \small E-mail: {\tt m.ortiz@ic.ac.uk}} \vskip 0.5truecm \today
\end{center}

\vskip 0.5truecm

\begin{abstract}
  We compute the canonical partition function of 2+1 dimensional 
  de Sitter space using the Euclidean $SU(2)\times SU(2)$ Chern-Simons
  formulation of 3d gravity with a positive
  cosmological constant. Firstly, we point out that one can work with
  a Chern-Simons theory with level $k=l/4G$, and its representations
  are therefore unitary for integer values of $k$. We then compute explicitly the partition function
  using the standard character formulae for $SU(2)$ WZW theory
  and find agreement, in the large $k$ limit, with the semiclassical result. Finally, we note that the de Sitter entropy can also be obtained as the degeneracy of states of representations of a Virasoro algebra with $c=3l/2G$. 
\end{abstract}

\medskip

\newpage

In 2+1 dimensions, general relativity has been shown to be equivalent
to Chern-Simons theory with a six dimensional gauge group whose
structure depends on the value of the cosmological constant and on the
signature of the spacetime metric \cite{AT-W}. In this letter, we
shall focus on the case of positive cosmological constant and
Euclidean metrics. This case has the attractive feature that the gauge
group is $SU(2)\times SU(2)$, and its representations are therefore
well defined.  We shall compute the canonical partition function of de
Sitter space by reducing the Chern-Simons theory to a boundary WZW
theory. For related calculations see \cite{ms,Park}. This calculation
is similar to those yielding the BTZ black hole entropy by considering
a boundary conformal field theory either at the black hole horizon
\cite{Carlip,bbo} or at spatial infinity \cite{Strom,ivo,bbo}.

The metric for 2+1 dimensional de Sitter space is
\begin{equation}
\label{dS}
ds^2 =  -\left(1 - \frac{r^2}{l^2}\right) dt^2 
       + \left(1 - \frac{r^2}{l^2}\right)^{-1} dr^2
       + r^2 d\phi^2
\end{equation}
where $\Lambda=1/l^2$ is the cosmological constant, and the horizon at
$r=l$ is that seen by an observer travelling along the timelike
geodesic at $r=0$. Performing a Wick rotation yields the Euclidean de
Sitter metric
\begin{equation}
\label{EDeS2}
ds^2_E =  \left(1 - \frac{r^2}{l^2}\right) dt_E^2 
       + \left(1 - \frac{r^2}{l^2}\right)^{-1} dr^2
       + r^2 d\phi^2
\end{equation}
where the period $\beta$ of the time coordinate $t_E$ ($0\leq t_E <\beta$) is fixed as $\beta =2\pi l$ by the requirement that the metric be everywhere regular.  Below, we shall allow a conical singularity at $r=0$. This will change the value of $\beta$ such that the metric is still regular at $r=l$.  

A simple coordinate transformation $r = l \cos \varrho$ makes it clear
that this is a three sphere, since
\begin{equation}
\label{EDeS3}
ds^2_E =  \sin^2\varrho \:  dt_E^2 + 
l^2 d\varrho^2 + l^2 \cos^2 \varrho \: d\phi^2.
\end{equation}
Note that $\varrho = 0$ describes the horizon and it is the origin of
the angular coordinate $t_E$, while the observer's worldline lies
at $\varrho = \pi/2$ and is the origin of $\phi$.  In the following, it will be convenient to introduce a new time coordinate $x^0$
with period 1 related to $t_E$ by $x^0 = t_E/\beta$.  

In computing the entropy of de Sitter space, we shall consider a
boundary surface that encloses the Euclidean world line of the
timelike geodesic observer.  (This surface may be thought of as
analogous to a surface at spatial infinity in the black hole case.)
The boundary surface results from removing (a thickened version of)
the observer's worldline from Euclidean de Sitter space. Since a three
sphere can be constructed by gluing together a pair of solid tori, the
remaining space is itself a solid torus.  The topology of the
Euclidean manifold that we consider is thus exactly the same as for
the BTZ black hole \cite{bbo}.

A gravity theory of Riemannian metrics with a positive cosmological
constant is classically equivalent to a Chern-Simons theory for the
group $SU(2)\times SU(2)$. The corresponding $SU(2)$ gauge fields are
\begin{equation}
^+A^a = \omega^a + {1\over l} e^a, \ \ \ ^-A^a = \omega^a - {1\over l}
e^a,
\end{equation}
where $w^a$ and $e^a$ are the spin connection and triad respectively. 

Using the metric \rref{EDeS3} we can read off the triad and spin
connection. Actually, we shall consider a generalization of the spacetime 
(\ref{EDeS3}) described by the two-parameter gauge field
\begin{eqnarray}
^{\pm}A^1 &=& -\gamma \sin \varrho(d\phi \mp \frac{\beta}{l} dx^0 ),
\nonumber \\
^{\pm}A^2 &=& \pm d\varrho, \nonumber \\
^{\pm}A^3 &=& \pm \gamma \cos \varrho (d\phi \mp \frac{\beta}{l} dx^0 )
\label{Egaugefield2}
\end{eqnarray}
from which de Sitter space is obtained for $\gamma=1$.  The parameter $\gamma$ parametrizes the deficit angle of a conical singularity located at $\varrho = \pi/2$ ($r=0$), or equivalently a holonomy around a non-contractible loop in the solid torus.  Indeed,
since the origin of $\phi$ is located at $\varrho=\pi/2$, and
$A^1_\phi = -\gamma$ at that point, the gauge field (\ref{Egaugefield2}) has a holonomy for $\gamma \neq 1$\footnote{Given an angular coordinate $\theta$ with period $2\pi$  whose origin is located at a point $r=0$, in general, there is a 
  holonomy around $r=0$ if the zero mode of $A_\theta$ is different
  from zero at $r=0$.  Suppose $A = \gamma d\theta J$ with $J$
  any of the hermitian $SU(2)$ generators. This value of $A$ can be
  set equal to zero by the gauge transformation $g=e^{-\gamma \theta
    J}$. However, $g$ is single valued only if $\gamma$ is an
  integer. In this case, the above holonomy is trivial. For
  $\gamma\neq n$, $g$ is multivalued and therefore $A$ does have a
  non-trivial holonomy. }.  We shall only consider the case $0<\gamma \leq 1$ which, from the metric point of view, represents an angular deficit in
$\phi$. We shall see below that the condition $\gamma<1$ will be
protected quantum mechanically by the bound in the spin ($2s< k$)
arising in affine $SU(2)$ representations.

Since $x^0$ is also an angular coordinate whose origin is located at $\varrho=0$, the requirement that there be no conical singularities or holonomies at the horizon ($\varrho = 0$) fixes $\gamma \beta/l = 2\pi $ and thus
\begin{equation}
\beta = { 2\pi l \over \gamma }.
\label{beta/g}
\end{equation}
For $\gamma=1$ we recover the period of de Sitter space and
(\ref{Egaugefield2}) yields the Euclidean metric (\ref{EDeS3}). 

The conical singularities arising in three dimensional gravity with a non-zero cosmological constant were introduced in \cite{Deser-Jackiw}. Their statistical mechanical properties have been recently studied in \cite{Park}.  Note that the presence of this singularity provides another reason to remove the observer's wordline $\varrho=\pi/2$ since the singularity is located at that point. The conformal field theory lives precisely on that surface. 

In the following we shall consider only the positive chirality gauge
field $^+A$ and denote it simply by $A$.  All the following equations
can be straightforwardly generalized to include the other $SU(2)$
field.

We work here in the canonical ensemble.  The first step in defining
the canonical partition function is to formulate boundary conditions.
Motivated by the gauge field (\ref{Egaugefield2}), and as in BTZ case
\cite{bbo,bg}, we fix the boundary conditions at $\varrho=\pi/2$ to be
\begin{equation}
\label{EDeS12}
A_0 = -\frac{\beta}{l} A_\phi, 
\end{equation}
where $\beta$  will be
the argument of the canonical partition function.  Since $\beta$
is the period of the Euclidean time coordinate, it can be 
interpreted as the inverse temperature. As explained before, de Sitter
space has $\beta=2\pi l$. Other values of $\beta$ correspond
to holonomies in the gauge field, and, for $\beta >2\pi l$, they can
be interpreted as conical singularities. 

As in \cite{bbo} we shall also add a term at the horizon $\varrho=0$
that imposes the constraint
\begin{equation}
\label{EDeS13}
A^{a}_0|_{\varrho=0} = - 2\pi \delta^a_3 
\end{equation}
which can be achieved by using a Wilson line along the horizon
\cite{bbo}.

The boundary conditions (\ref{EDeS12}) and (\ref{EDeS13}) are
motivated by the on-shell gauge field (\ref{Egaugefield2}) associated
to de Sitter space. However, they give rise to a far bigger space.
Indeed, the phase space is infinite dimensional and is described by a
chiral WZW model (see below).

The Euclidean action appropriate to our boundary conditions
((\ref{EDeS13}) and (\ref{EDeS12})) is \footnote{We use the
  conventions $T_a = -(i/2)\sigma_a$, $[T_a,T_b]= -\epsilon_{abc}
  T^c$, $Tr(T_aT_b) = -\frac{1}{2} \delta_{ab}$. The total action
  including both chiralities is $I = I[^+A,^+\beta] - I[^-A,^-\beta]$.
  }
\begin{eqnarray}
\label{EDeS15}
I_E[A,\beta] &=& \frac{k}{4\pi} \int_M \epsilon^{kl} \mbox{Tr} \left(iA_k
\partial_0 {A}_l - A_0 F_{kl} \right) d^2xdx^0 \\
  && -\frac{k\beta}{4l\pi} \int_{\varrho=\pi/2} \mbox{Tr} (A_\phi)^2 d\phi dx^0 - \frac{k}{2} \int_{\varrho=0} A_\phi^{(3)} d\phi dx^0. \nonumber 
\end{eqnarray}
with $k = l/4G$. 

This choice of action requires some explanation. The Chern-Simons action does not depend on the metric and therefore the spacetime signature does not affect its form.  In the Chern-Simons formulation of General Relativity, the information about the spacetime signature is contained in the local gauge group.

However, let $k$ be a positive integer, ${\cal G}$ a compact Lie group, and consider the Chern-Simons action $I[A]= (k/4\pi) \int_M \mbox{Tr} \epsilon^{ij}(-A_i \dot A_i + A_0 F_{ij})\ $ (plus boundary terms) on a manifold $M$ with the topology $\Sigma \times S_1$.  We take $S_1$ as the time direction and, as usual in Euclidean field theory, we relate its period with the inverse temperature $\beta$.  We now ask the question of what is the right measure in the path integral which produces a partition function of the form Tr $e^{-\beta H}$, with $H$ real and positive, and the trace being taken over a well defined Hilbert space. If we integrate $e^{iI[A]}$, one finds a Kac-Moody algebra at level $k$ whose representations are well defined, but the Hamiltonian has an unwanted $i\beta$ coefficient. If, on the contrary, we integrate $e^{-I[A]}$, one finds the right measure $e^{-\beta H}$, but the representations are not well defined because this yields a Kac-Moody algebra at level $ik$.  The source of this problem is not the choice of the measure but the action itself. Indeed, Chern-Simons theory is of first order (linear in the time derivatives) therefore its correct ``Euclidean" version is $I_E[A] = (k/4\pi) \int \mbox{Tr} \epsilon^{ij}(i A_i \dot A_i + A_0 F_{ij})$ (plus boundary terms) and one integrates $e^{-I_E}$.  This is the form of the action (\ref{EDeS15}).  We stress that here the terminology ``Euclidean" refers to properties of the path integral and not to the spacetime signature (which is encoded in the local gauge group).   

To further justify (\ref{EDeS15}) we note, first, that the equations of motion derived from it are $F_{ij}=0$ and $i\partial_0 {A}_i = D_i A_0$ and they are solved by the de Sitter gauge field (\ref{Egaugefield2}), and second, the value of $I$ on the solution (\ref{Egaugefield2}) coincides with that of the Euclidean Hilbert action on the metric (\ref{EDeS2}) with
$k=l/4G$ (see below).  Finally, and perhaps most importantly in the context of Euclidean integrals in quantum gravity, the action (\ref{EDeS15})
gives rise to a well defined partition function that can be computed
exactly and yields sensible answers. In a future publication
\cite{bbo3}, we shall argue that a similar approach leads to a
simplified treatment of Riemannian anti-de Sitter metrics, although in
that case the gauge group is necessarily complex.

The partition function associated to $A$ is then equal to 
\begin{equation}
Z_A(\beta) = \int D[A] \exp\left(-I_E[A,\beta] \right).
\label{Z}
\end{equation}
where the measure $D[A]$ denotes a sum over all gauge fields modulo
gauge transformations and $I[A,\beta]$ is given in (\ref{EDeS15}).
Note that this partition function has the right semiclassical value.
Indeed, in the saddle point approximation provided by the de Sitter
gauge field (\ref{Egaugefield2}) one finds
\begin{equation}
Z_A(\beta) \sim \sum_\gamma e^{-\beta \gamma^2/(16 G)+\pi \gamma l/(4G) } .
\label{Zr}
\end{equation}
The sum over $\gamma$ arises because the only fixed quantity in the
canonical calculation is $\beta$. A saddle point approximation for
that sum yields the relation $\gamma \beta = 2\pi l$ which is the
classical value of $\gamma$ that avoids conical singularities at the horizon.  From (\ref{Zr}) we can associate to each
state labeled by $\gamma$ an energy $E_\gamma = \gamma^2/(16G)$ and a
degeneracy $\rho(\gamma)=\exp (\pi \gamma l/(4G))$. The de Sitter
state corresponds to $\gamma=1$ and has a total entropy (adding the
entropy coming from the other $SU(2)$ gauge field) $S=S_l + S_r = \pi
l/(2 G)$ which is the Gibbons-Hawking value for de Sitter space
\cite{GH}.

Our goal now is to compute the above partition function exactly. In
the exact calculation we shall re-encounter the sum (\ref{Zr}) but
with discretized values of $\gamma$. The condition $\gamma<1$, which
is necessary for a sensible metric interpretation for the sum over
$\gamma$ as conical singularities, will be protected by the bound on
the spin $s$ arising in affine $SU(2)$ representations.

It is well known that (\ref{Z}) can be reduced to a $WZW$ model
at the boundary. The idea is that once the bulk gauge degrees of
freedom are eliminated, one is left with an effective theory
describing an infinite dimensional residual gauge group at the
boundary. This infinite dimensional group is parametrized by a group
element $g$ and the action is the chiral WZW model\footnote{The
  reduction from Chern-Simons to $WZW$ has been extensively studied in
  the literature. In our situation, the simplest way to pass from
  (\ref{Z}) to (\ref{EDeS16}) is by solving the constraint $F_{kl}=0$. Alternatively, one can integrate over purely imaginary values of $A_0$.}
\begin{eqnarray}
\label{EDeS19}
I_{CWZW}[g,\beta] &=& -\frac{ik}{4\pi} \int  \mbox{Tr} (\partial_\phi g^{-1} \partial_\tau{g} )d\phi d\tau - \frac{ik}{12\pi} \int_M
     \mbox{Tr} (g^{-1}dg)^3 d^2xd\tau \\
 && -\frac{k \beta}{4 l\pi} \int  \mbox{Tr} (g^{-1}\partial_\phi g)^2d\phi d\tau 
    - \frac{k}{2} \int (g^{-1}\partial_\phi g)^{(3)} d\phi d\tau. \nonumber 
\end{eqnarray}  
The partition function thus becomes
\begin{equation}
\label{EDeS16}
Z(\beta) =  \int D[g] \exp \left(-I_{CWZW}[g,\beta]
           \right)
\end{equation}
and can be evaluated using canonical methods. The canonical
Poisson brackets associated to the $WZW$ action are given by the
$SU(2)$ Kac-Moody algebra,
\begin{equation}
\label{EDeS20}
[T^a_n,T^b_m]= i{\epsilon^{ab}}_c T^c_{n+m} + n \frac{k}{2}
              \delta^{ab} \delta_{n+m,0},
\end{equation}
where the $T^a_n$ are the Fourier components of the gauge field,
\begin{equation}
\label{EDeS21}
A_\phi = g^{-1} \partial_\phi g = \frac{2}{k}
\sum_{n=-\infty}^{\infty}
         T_n^a e^{in\phi}.
\end{equation}    
$Z$ can thus be computed as
\begin{equation}
\label{EDeS22}
Z_A(\beta) = \sum_{2 s=0}^k {\rm Tr}_{s} \: 
          \exp \left(-\beta L_0/l
           + 2\pi  T_0^3 \right),
\end{equation}
where $s$ labels the spin of the different $SU(2)$ representations and
$L_0$ is a Virasoro generator defined by
\begin{equation}
\label{EDeS23}
L_0 = \frac{1}{k+Q} \sum_{n=-\infty}^{\infty} :T^a_{-n} T^b_{n}:
\delta_{ab}.
\end{equation}
The parameter $Q$ is the second Casimir in the adjoint representation.
We shall be interested in the $k \gg Q$ limit so this term can be
neglected.  The computation of the trace in \rref{EDeS22} follows from
\cite{Goddard-Kent-Olive},
\begin{eqnarray}
\mbox{Tr}_s \left(q^{L_0}e^{i \theta
T_0^3}\right) 
&=& \frac{ q^{\frac{s(s+1)}{k}}  \sum_{n=-\infty}^\infty 
                q^{k n^2 + (2s +1) n} 
                \left\{ e^{i(s+k n)\theta}
                -e^{-i(s+1+k n)\theta }\right\} }{ 
                \Pi_{m=1}^\infty (1-q^m)(1-q^m e^{i\theta})
                          (1-q^{m-1}e^{-i\theta})} \nonumber \\ 
&=&  q^{\frac{s(s+1)}{k}} D^{-1}  \sum_{n=-\infty}^\infty 
                q^{k n^2 + (2s +1) n} 
                \frac{\sin\left[\left(s+k n+
                 \frac{1}{2}\right)\theta \right]}{ 
                \sin{\frac{\theta}{2}}} 
\label{chGKO}
\end{eqnarray}
with $q=e^{-\beta/l}$ and $D= \Pi_{m=1}^\infty (1-q^m)(1-q^m e^{i\theta})
(1-q^m e^{-i\theta})$. Splitting the sine in \rref{chGKO} according to
$\sin(x+y) = \sin(x) \cos(y) + \cos(x) \sin(y)$, using the fact that 
in our case $\theta= -2 \pi i$, and assuming the main contribution in the
trace over the spin representations to come from high values of $s$,  
one finally ends up with an effective 
partition function 
\beq{chGKO4}
Z_A(\beta) = \sum_{2s=0}^{k} D^{-1}
     q^{\frac{s(s+1)}{k}}  \frac{\sinh\left[2\pi (s+\frac{1}{2}) 
             \right]}{\sinh \pi} 
        \sum_{n=-\infty}^\infty 
                q^{k n^2 + (2s +1) n} 
                                                e^{2 \pi k n }.
\eeq
In the  semiclassical limit $k\rightarrow \infty$ we can concentrate
on the numerator since it carries all the $k$ dependence whereas
the denominator leads to a subleading contribution
\footnote{For the denominator 
$D = \Pi_{n=1}^{\infty} (1-q^n)(1-q^n e^{2\pi})(1-q^n e^{-2\pi})$
we  get a subleading contribution to the density of states, but
which interestingly is equal to zero for de Sitter space $-\beta/l=\log
q=-2\pi$. But this is only the case if we neglect the fact that   
the temperature gets renormalised in the quantum 
calculation. This is because the particular form of the quantum Virasoro 
operator $L_0$ defind in (\ref{EDeS23}) forces us to replace 
$\beta$ by $\beta' = (1+ Q/k) \beta$.}.  

The sum over $n$ can be calculated by a saddle point approximation
obtaining for the saddle point
\begin{equation}
n = {\gamma \over 2} - {1 \over 2k} - {s \over k}
\label{n}
\end{equation}
where we have replaced $\beta = 2\pi l /\gamma$. Since $s$ is
bounded from above by $k/2$ and $\gamma<1$, in the limit $k
\rightarrow \infty$ we have $n <1$ and thus the sum is well approximated by
$n=0$. 

We thus find an effective quantum mechanical partition function
 \begin{equation}
Z_A[\beta] = \sum_{2s=0}^{k}  e^{2\pi s}\, e^{-\beta
s(s+1)/lk }
\label{Ze}
\end{equation}
which should be compared with the semiclassical sum (\ref{Zr}). 
The energy levels are quantized and given by  
\begin{equation}
E_s  =  {s(s+1) \over lk},
\label{E}
\end{equation}
with apparent degeneracies $\rho(s) = e^{2\pi s}$.  We calculate the sum (\ref{Ze}) by a saddle point approximation. The saddle point occurs when $\beta$ and $s$ are related as
\begin{equation}
\beta = 2\pi l \frac{k}{(2s+1)}.
\label{beta}
\end{equation}
As we mentioned above, the values of $\beta$ of
the form $\beta = 2\pi l/\gamma$ with $\gamma<1$ can be interpreted as
conical singularities with an angular deficit defined by $\gamma$. Thus
we find the series of quantized conical singularities with
\begin{equation}
\gamma = {2s+1 \over k} 
\end{equation}
which is indeed less than 1 for the allowed states $0\leq 2s \leq
k$.  

The state with $s$ taking its maximum value $s=k/2$ has 
$\beta = 2\pi l$ (in the limit $k \rightarrow \infty$) and corresponds to de Sitter space. Its energy, according to (\ref{E}), is $E = 1/16G$ and degeneracy is $e^{\pi k}$.  Adding the degeneracy associated to the other $SU(2)$ gauge field (whose de Sitter state has the same degeneracy) yields the total entropy
\begin{equation}
S_{dS} = {2\pi k} = \frac{2\pi l}{4 G}
\end{equation}
which agrees with the semiclassical approximation for the
Gibbons-Hawking entropy of de Sitter space.

Note, that in analogy with the BTZ black hole \cite{bbo} the density of 
states arises purely from the presence of the horizon term. This
raises the question of whether the density of states could arise
dynamically in the context of a theory that does
not presuppose the relation (\ref{EDeS13}).

Finally, we point out an interesting relation between this calculation
and the use of the twisted Sugawara construction to produce a pair
of Virasoro algebras with a classical central charge. This discussion
follows the ideas developed in \cite{CHvD,max} but without making a direct
connection between the Virasoro algebras and the group of symmetries
of de Sitter space at the boundary.

The gauge field \rref{Egaugefield2} obeys the boundary conditions set
out in Refs. \cite{bbo,max} that allow us to define a pair of Virasoro
algebras at the boundary. The central charge of the algebra is
\begin{equation}
\label{DeS8}
c = 6 k = \frac{3 l}{2 G}
\end{equation}
as in the case of negative cosmological constant. This central charge
is fixed in the case of negative cosmological constant by the
condition that the Virasoro charges leave invariant the anti-de Sitter
metric at infinity \cite{BH,CHvD,bbo,mm}. In the present case the only
justification we have for this choice is by analytic continuation of
the result of \cite{bbo} and that it turns out to give the correct
result. Another difference is that in this case our algebra comes from
a theory of Euclidean rather than Lorentzian metrics, since this
allows us to avoid problems of complex gauge group and complex $k$.

The Virasoro operators $L_0$ and $\bar{L}_0$ are defined as
\begin{equation}
\label{DeS9}
L_0 = -\frac{k}{4\pi} \int Tr \left( A^2 -{1\over 2} \right)d\phi =
\frac{l}{8G}
\end{equation}
\begin{equation}
\label{DeS10}
\bar{L}_0 = -\frac{k}{4\pi} \int Tr \left( \tilde{A}^2 -{1\over 2}
\right)d\phi = 
    \frac{l}{8G}
\end{equation}
and from this we see that de Sitter space can be defined by the
condition that the asymptotic zero modes satisfy the conditions
\begin{equation}
L_0 = c/12,\qquad \bar{L}_0 = c/12.
\label{zero}
\end{equation}
Note that, just as in the anti-de Sitter case, the Virasoro operators
which yield the central charge normalized as $(c/12) n(n^2-1)$ are
shifted with respect to the mass by $M=(L_0-c/24)+(\bar L_0-c/24)$.
The de Sitter energy of each sector is then $lE_{dS} = c/24 = l/16G$,
just as in the canonical calculation.  

It is now interesting to see whether one can extract the density of
states from the Cardy formula as in the black hole case
\cite{bbo,Strom,ivo}. However, the formula used in those papers
is only valid if the eigenvalue $N$ of $L_0$ is much larger than $c$.
Eq.  \rref{zero} shows that this is not the case.  
A generalization of the formula is straightforward and can be 
derived from \cite{Carlipl}. 
The improved formula reads\footnote{Note 
that the use of the formula \rref{DeS11} implies the assumption
(that we have as yet been unable to verify) 
that $Z(-\frac{1}{\beta})$ is a slowly varying function in the region
of the saddle point.}
\begin{equation}
\label{DeS11}
\varrho(l) = \exp\left[ 2\pi \sqrt{\frac{c}{6}  
                  \left(N - \frac{c}{24} \right)}
              +  2\pi \sqrt{\frac{\bar{c}}{6} \left(\bar{N} 
                 - \frac{\bar{c}}{24}  \right)}\right]
\end{equation}
which reduces to the formula used in \cite{bbo} in the case of large
$N$.
The entropy following from this formula is 
\begin{equation}
\label{DeS12}
S = \frac{\pi l}{2 G}
\end{equation}
in exact agreement with the standard result.

This last result at least suggests that a theory in which the density
of states arises dynamically is likely to have an effective central
charge of $3l/2G$ as in Anti-de Sitter space.

\section*{Acknowledgements}
We are grateful to Fay Dowker, Marc Henneaux and Adam Ritz for helpful conversations. MB was supported by CICYT (Spain) project AEN-97-1680
and also thanks the Spanish postdoctal program of Ministerio de Educaci\'on y Cultura.  TB acknowledges financial support from the
German Academic Exchange Service (DAAD). MEO was supported by the
PPARC, UK.

\end{document}